\documentclass[a4paper,11pt]{article}
\usepackage{pos}
\usepackage{outlines}
\usepackage{graphicx}
\usepackage{hyperref}
\usepackage{lineno}
\usepackage{placeins}

\title{Searching for Sources of High Energy Neutrinos from Magnetars with IceCube}

\author{The IceCube Collaboration \\{\normalsize \normalfont(a complete list of authors can be found at the end of the proceedings)}}

\emailAdd{ava.ghadimi@icecube.wisc.edu}
\emailAdd{jmsantander@ua.edu}

\abstract{Magnetars are neutron stars with very strong magnetic fields on the order of $10^{13}$ to $10^{15}$ G. Young magnetars with oppositely-oriented magnetic fields and spin moments may emit high-energy (HE) neutrinos from their polar caps as they may be able to accelerate cosmic rays to above the photomeson threshold \cite{Zhang_2003}. Giant flares of soft gamma-ray repeaters (a subclass of magnetars) may also produce HE neutrinos and therefore a HE neutrino flux from this class is potentially detectable \cite{Ioka_2005}. Here we present plans to search for neutrino emission from magnetars listed in the McGill Online Magnetar Catalog using 10 years of well-reconstructed IceCube muon-neutrino events looking for significant clustering around magnetars' direction. IceCube is a cubic kilometer neutrino observatory at the South Pole and has been fully operational for the past ten years.

\vspace{4mm}
{\bfseries Corresponding authors:}
Ava Ghadimi$^{1*}$, Marcos Santander$^{1}$\\
{$^{1}$ \itshape University of Alabama}\\
\\[4mm]
$^*$ Presenter

\FullConference{37$^{\rm{th}}$ International Cosmic Ray Conference (ICRC 2021)\\
		July 12th -- 23rd, 2021\\
		Online -- Berlin, Germany}

}

\begin{document}
\maketitle

\section{Introduction}\label{sec:intro}

Magnetars are neutron stars with very strong magnetic fields on the order of $10^{13}$ to $10^{15}$ G. Magnetars are classified into two groups: Anomalous X-ray Pulsars (AXPs), and Soft Gamma-ray Repeaters (SGRs). They are strong x-ray emitters and in their early stages maintain high x-ray luminosity over a long period of time \cite{Zhang_2003}.

Young magnetars with oppositely oriented magnetic fields and spin moments may emit high-energy (HE) neutrinos from their polar caps as they may be able to accelerate cosmic rays to above the photomeson threshold. Since their x-ray luminosity spans a long period, they should contribute higher neutrino flux than older magnetars. Young magnetars may also contribute to the astrophysical diffuse neutrino background \cite{Zhang_2003}.

Post-burst magnetars (SGRs after flaring) show an increase in their quiescent luminosity for a long period of time, therefore they should contribute higher neutrino fluxes \cite{Zhang_2003}.

 Giant flares of SGRs may produce HE neutrinos and therefore a HE neutrino flux from this class is potentially detectable by IceCube \cite{Ioka_2005}. They hypothesize that "... the baryon-rich model with a flare $10^{-3}$ times smaller than that considered here [SGR 1806-20] can produce about one event in IceCube, and the rate of such flares is about $\thicksim 1/10$ year".

\section{Motivation}

\subsection{Neutrino Emission Mechanism in Magnetars}

Neutrino emission due to the acceleration of high-energy protons usually happens through pion decay, where the protons interact with the photons or matter in the environment of astrophysical accelerators. For the case of most pulsars, the immediate environment such as the magnetosphere lacks a target column density large enough for pion production. Therefore, the neutrino emission process in pulsars is usually considered to happen in the pulsar wind nebulae \cite{Beall_2002}.
However, in the inner magnetosphere of pulsars with surface magnetic fields of $\thicksim 10^{15}$ G, i.e. magnetars, conditions for neutrino production via photomeson interaction are realized. 

In principle, there are two main sources of energy that power a magnetar: the spin-down power, and the power resulting from decaying magnetic fields (magnetic power). The spin-down power accelerates protons and the magnetic power provides a large amount of near-surface photons. Assuming both of these energy sources power the magnetar, and that the magnetar is young enough, then the criterion for photomeson interactions are satisfied \cite{Zhang_2003}.

The dominant photomeson interaction resulting in neutrino emission in magnetars then is through the $\Delta$-resonance \cite{Zhang_2003}:

\begin{equation}
    p\gamma \rightarrow \Delta \rightarrow n\pi^{+} \rightarrow n\nu_{\mu}\mu^{+} \rightarrow n\nu_{\mu}e^{+}\nu_{e}\bar{\nu_{\mu}}
\end{equation}

\subsection{Neutrino Detection: the IceCube Neutrino Observatory}

The IceCube Neutrino Observatory is a cubic-kilometer neutrino detector deep inside the Antarctic ice \cite{Aartsen_2017} and has been operational for the past 10 years. In 2013, IceCube published the first evidence of HE neutrinos of astrophysical origins \cite{1242856}. 

As the HE neutrinos interact with the Antarctic ice, they produce relativistic charged particles which emit Cherenkov light. The detector consists of 5160 digital optical modules (DOMs) which can detect the Cherenkov light. Using the signals from the DOMs, one can infer the energy, direction, and flavor of the HE neutrino.

The charged-current interactions of muon neutrinos produce high-energy muons that can travel kilometers in the ice. These muon tracks have an angular resolution of $\thicksim 1^{\circ}$ for energies above 10 TeV. In this work we plan to use a sample of events from both the northern and southern sky using 10 years of IceCube data which are optimized for searches for astrophysical neutrino point sources. \cite{PhysRevLett.124.051103}.

\section{Analysis Plan}

In this work, we will use the McGill Online Magnetar Catalog \cite{Olausen_2014}. The catalog consists of 30 magnetars, 16 SGRs and 14 AXPs, the majority of which are in the Southern sky and within $\thicksim 60$ kpc. 3 magnetars are off-plane: one is spatially coincident with the Large Magellanic Cloud (LMC), one is spatially coincident with the Small Magellanic Cloud (SMC), and one is located in the Northern sky outside of the Galactic plane. The catalog contains the persistent characteristics of the magnetars such as period, x-ray flux, magnetic field strength, etc, unless otherwise noted. 

\begin{figure}[!h]
    \centering
    \includegraphics[width=1\textwidth]{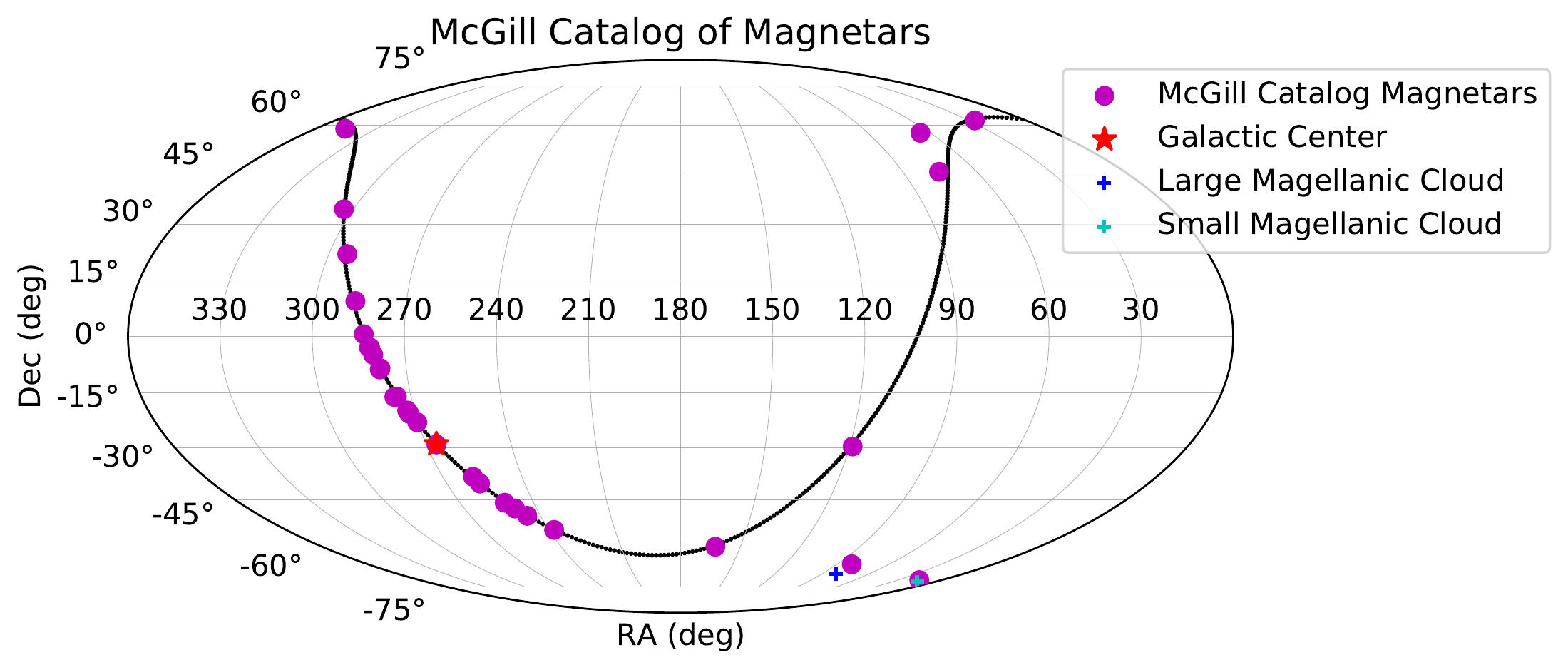}
    \caption{Location of the magnetars in the McGill catalog in Equatorial coordinates. One magnetar is spatially coincident with LMC, and one is spatially coincident with SMC.}
\end{figure}

In addition to the McGill catalog, the following magnetars will be included in this work:
The two newly discovered magnetars, Swift J1555.2-5402 and SGR 1830-0645 will be included in all 3 phases discussed below.
$\gamma$-ray burst GRB 200415A is believed to be a giant flare of a magnetar in the starburst galaxy NGC 253 \cite{Svinkin2021}. We will include NGC 253 in our time-dependent (section 3.2) and individual analyses (section 3.3).

\newpage
There are three phases in our search for neutrinos from magnetars with IceCube:

\subsection{Time-integrated Stacked Search}

We plan to use a point source search using the unbinned Likelihood method. We will perform a time-integrated stacked analysis to increase the sensitivity.

The likelihood function for a single point source is given by: 
\begin{equation}
    \mathcal{L}(x_s,n_s) = \prod_{i = 1}^{N}  \left(\frac{n_s}{N}\mathcal{S}(x_i,x_s,E_i,\gamma) +\left(1 -\frac{n_s}{N}\right)\mathcal{B}(x_i, E_i)\right)
\end{equation}

Where $x_s$ is the position of the source, $n_s$ is the number of the signal events, $x_i$ and $E_i$ are the position and energy of the $i$th neutrino candidate event (hereafter event), N is the total number of events, and $\gamma$ is the spectral index which is fit globally across all sources.

The background PDF $\mathcal{B}$ is a function of the reconstructed energy and the declination of the events. The background PDF does not depend on the right ascension (RA) since the effective area of the detector, averaged over time, is constant with respect to RA.

The signal PDF  $\mathcal{S}$ is assumed to only have spatial and energy components for the time-integrated search and to be Gaussian in form.

The weights used in the analysis are as follows:
\begin{itemize}
    \item \textbf{Equal}: Probe magnetars as a general class without taking into account any models.

    \item \textbf{Energy flux}: Neutrino flux and the unabsorbed X-ray energy flux have a direct correlation.

    \item \textbf{Inverse Period}: Young magnetars are more likely to emit high energy neutrinos. Characteristic age of the magnetar depends directly on the period.
\end{itemize}

Therefore, the likelihood function for the stacked analysis is given by:

\begin{equation}
    \mathcal{L}(x_s,n_s) = \prod_{i = 1}^{N}  \left(\sum_{j=1}^{M}\frac{n_s}{N}w_j\mathcal{S}_{j}(x_i,x_s,E_i,\gamma_j) +\left(1 -\frac{n_s}{N}\right)\mathcal{B}(x_i, E_i)\right)
\end{equation}

where the index $j$ denotes the source from our catalog, and $w_j$ is the weight.

\subsubsection{Testing the Neutrino Emission Model in Zhang, et al. \cite{Zhang_2003}}

\begin{figure}[!h]
\centering
\includegraphics[width=0.9\textwidth]{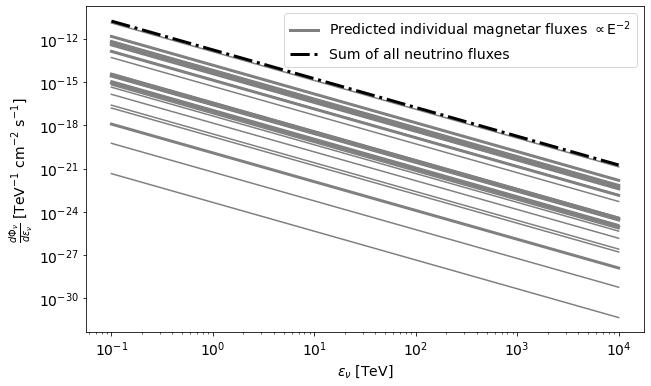}
\caption{Neutrino fluxes according to Zhang et al. \cite{Zhang_2003} using the data available in the McGill Online Magnetar catalog \cite{Olausen_2014}}
\label{Zhang nu flux}
\end{figure}

The neutrino number flux arriving at Earth is given by equation 15 in \cite{Zhang_2003}.
Taking the derivative $\frac{d\phi_{\nu}}{d\epsilon_{\nu}}$ gives us the differential neutrino flux from the magnetars. Using the data in the McGill magnetar catalog \cite{Olausen_2014}, we have plotted the differential neutrino flux in Figure \ref{Zhang nu flux}. We will compare these to the sensitivity and discovery potentials obtained from our time-integrated stacking analysis. 

\FloatBarrier

\subsection{Time-dependent Search}

We will test the hypothesis that the baryon-rich flare of a SGR $10^{-3}$ times smaller than that of SGR 1086-20 can produce about one event in IceCube, and the rate of such flares is about $\thicksim 1/10$ year \cite{Ioka_2005}. To do this, we will perform a time-dependent light curve analysis using the method in \cite{2020arXiv201201079I}.\\

\begin{figure}
    \centering
    \includegraphics[width=0.9\textwidth]{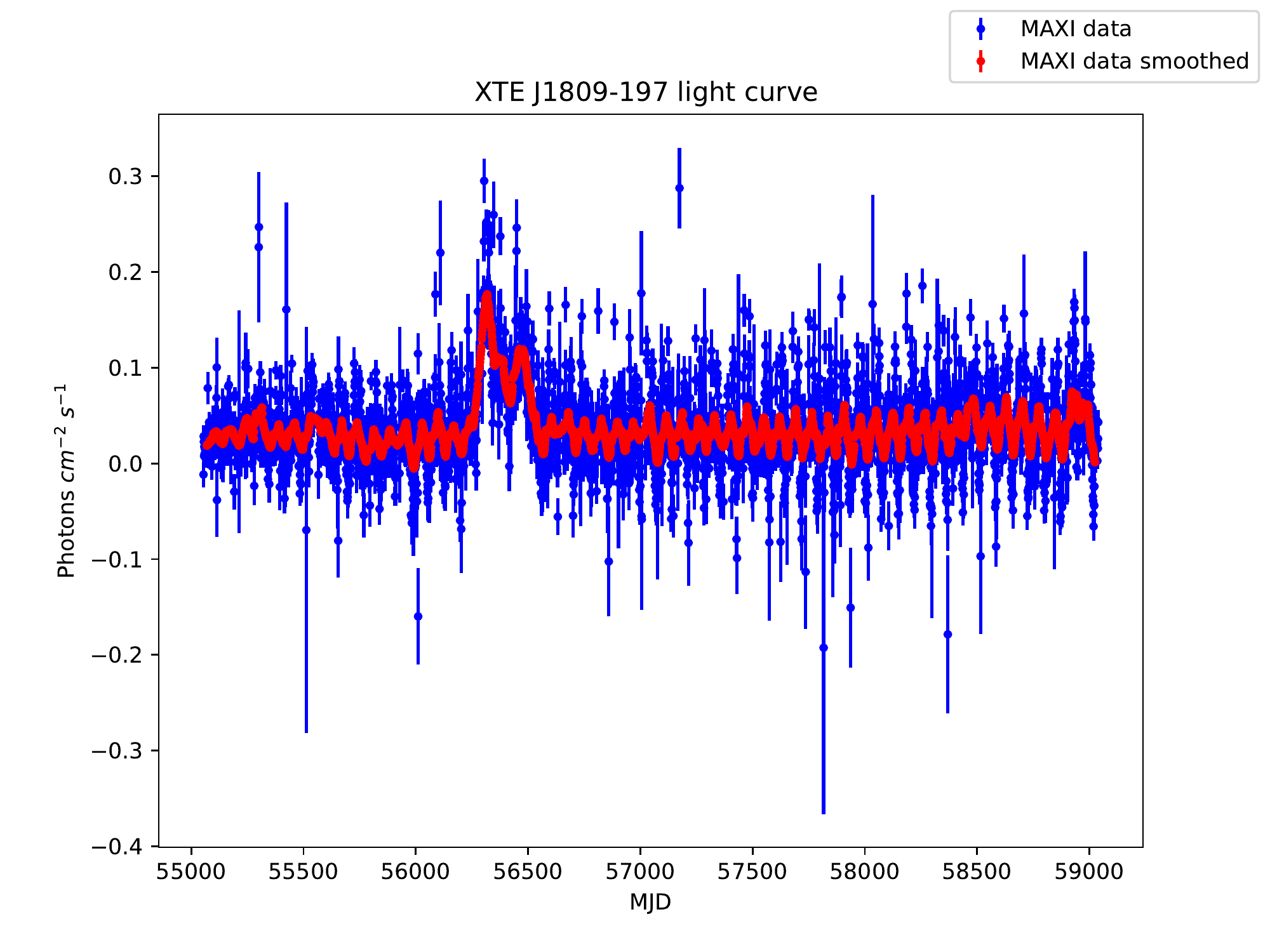}
      \includegraphics[width=0.9\textwidth]{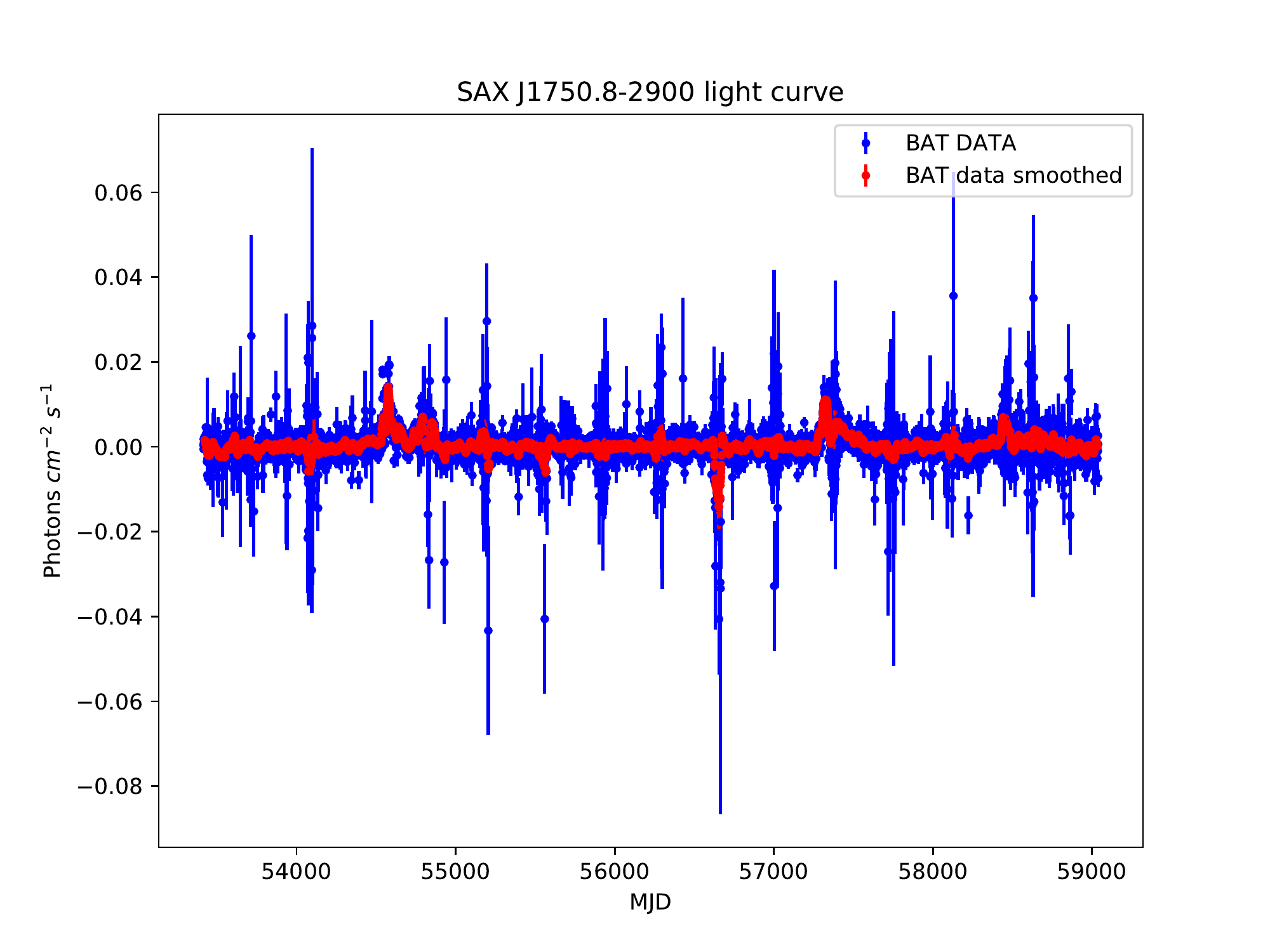}

    \caption{Top: X-ray light curves of XTE J1809-197 from  \href{http://maxi.riken.jp/top/lc.html}{MAXI/Riken}.
    Bottom: SAX J1750.8-2900 from  \href{https://swift.gsfc.nasa.gov/results/transients/BAT_current.html}{Swift-BAT}. Both light curves show possible variability.}
    \label{light curves}
\end{figure}

To justify the time-dependent analysis, we looked for variability in the X-ray light curve of the magnetars in our catalog using the data from MAXI/Riken and Swift BAT. Figure \ref{light curves} shows the light curve of two magnetars which exhibit potential variability in their light curve. We will search for neutrinos in IceCube around the time of increased X-ray activity of the magnetars. 

\subsection{Study of Individual Sources}

Lastly, we will look at individual sources such as SGR 1935+2154, which is associated with a fast radio burst (FRB) \cite{Bochenek2020}, without stacking to set upper limits on the neutrino flux.

\section{Summary and Future Work}

Here we presented a proposed search for neutrino emission from magnetars using 10 years of IceCube data. We also outlined the three phases of this analysis and discussed the rationale behind each. We are now in the process of generating sensitivities and discovery potentials for the time-integrated stacking analysis. In case no significant signal is identified, we will then set constraints on magnetars as a whole class of objects contributing to the all-sky astrophysical neutrino flux. Following the time-integrated stacking analysis, we will move on to probing magnetars as transient sources by performing a time-dependent analysis. Finally we will look at individual magnetars that are of special interest, such as SGR 1935+2154 which has been associated with a FRB \cite{Bochenek2020}.

\bibliographystyle{ICRC}
\bibliography{references}



\clearpage
\section*{Full Author List: IceCube Collaboration}




\scriptsize
\noindent
R. Abbasi$^{17}$,
M. Ackermann$^{59}$,
J. Adams$^{18}$,
J. A. Aguilar$^{12}$,
M. Ahlers$^{22}$,
M. Ahrens$^{50}$,
C. Alispach$^{28}$,
A. A. Alves Jr.$^{31}$,
N. M. Amin$^{42}$,
R. An$^{14}$
% ...
\\

\noindent
% ...
$^{12}$ Universit{\'e} Libre de Bruxelles, Science Faculty CP230, B-1050 Brussels, Belgium \\
% ...
$^{14}$ Department of Physics and Laboratory for Particle Physics and Cosmology, Harvard University, Cambridge, MA 02138, USA \\
% ...
$^{17}$ Department of Physics, Loyola University Chicago, Chicago, IL 60660, USA \\
$^{18}$ Dept. of Physics and Astronomy, University of Canterbury, Private Bag 4800, Christchurch, New Zealand \\
% ...
$^{22}$ Niels Bohr Institute, University of Copenhagen, DK-2100 Copenhagen, Denmark \\
% ...
$^{28}$ D{\'e}partement de physique nucl{\'e}aire et corpusculaire, Universit{\'e} de Gen{\`e}ve, CH-1211 Gen{\`e}ve, Switzerland \\
% ...
$^{31}$ Karlsruhe Institute of Technology, Institute for Astroparticle Physics, D-76021 Karlsruhe, Germany  \\
% ...
$^{42}$ Bartol Research Institute and Dept. of Physics and Astronomy, University of Delaware, Newark, DE 19716, USA \\
% ...
$^{50}$ Oskar Klein Centre and Dept. of Physics, Stockholm University, SE-10691 Stockholm, Sweden \\
% ...
$^{59}$ DESY, D-15738 Zeuthen, Germany \\
% ...

\subsection*{Acknowledgements}

\noindent
USA {\textendash} U.S. National Science Foundation-Office of Polar Programs,
U.S. National Science Foundation-Physics Division,
U.S. National Science Foundation-EPSCoR,
Wisconsin Alumni Research Foundation,
Center for High Throughput Computing (CHTC) at the University of Wisconsin{\textendash}Madison,
Open Science Grid (OSG),
Extreme Science and Engineering Discovery Environment (XSEDE),
Frontera computing project at the Texas Advanced Computing Center,
U.S. Department of Energy-National Energy Research Scientific Computing Center,
Particle astrophysics research computing center at the University of Maryland,
Institute for Cyber-Enabled Research at Michigan State University,
and Astroparticle physics computational facility at Marquette University;
Belgium {\textendash} Funds for Scientific Research (FRS-FNRS and FWO),
FWO Odysseus and Big Science programmes,
and Belgian Federal Science Policy Office (Belspo);
Germany {\textendash} Bundesministerium f{\"u}r Bildung und Forschung (BMBF),
Deutsche Forschungsgemeinschaft (DFG),
Helmholtz Alliance for Astroparticle Physics (HAP),
Initiative and Networking Fund of the Helmholtz Association,
Deutsches Elektronen Synchrotron (DESY),
and High Performance Computing cluster of the RWTH Aachen;
Sweden {\textendash} Swedish Research Council,
Swedish Polar Research Secretariat,
Swedish National Infrastructure for Computing (SNIC),
and Knut and Alice Wallenberg Foundation;
Australia {\textendash} Australian Research Council;
Canada {\textendash} Natural Sciences and Engineering Research Council of Canada,
Calcul Qu{\'e}bec, Compute Ontario, Canada Foundation for Innovation, WestGrid, and Compute Canada;
Denmark {\textendash} Villum Fonden and Carlsberg Foundation;
New Zealand {\textendash} Marsden Fund;
Japan {\textendash} Japan Society for Promotion of Science (JSPS)
and Institute for Global Prominent Research (IGPR) of Chiba University;
Korea {\textendash} National Research Foundation of Korea (NRF);
Switzerland {\textendash} Swiss National Science Foundation (SNSF);
United Kingdom {\textendash} Department of Physics, University of Oxford.
\end{document}